# Improving the Load Balancing Performance of Vlasiator

Ata Turk[a], Cevdet Aykanat[a], G. Vehbi Demirci[a], Sebastian von Alfthan[b], Ilja Honkonen[b]

[a]*Bilkent University, Computer Engineering Department, 06800 Ankara, Turkey*
[b] *Finnish Meteorological Institute, PO Box 503, Helsinki, FI-00101, Finland*

**Abstract**

This whitepaper describes the load-balancing performance issues that are observed and tackled during the petascaling of the Vlasiator codes. Vlasiator is a Vlasov-hybrid simulation code developed in Finnish Meteorological Institute (FMI). Vlasiator models the communications associated with the spatial grid operated on as a hypergraph and partitions the grid using the parallel hypergraph partitioning scheme (PHG) of the Zoltan partitioning framework. The result of partitioning determines the distribution of grid cells to processors. It is observed that the partitioning phase takes a substantial percentage of the overall computation time. Alternative (graph-partitioning-based) schemes that perform almost as well as the hypergraph partitioning scheme and that require less preprocessing overhead and better balance are proposed and investigated. A comparison in terms of effect on running time, preprocessing overhead and load-balancing quality of Zoltan's PHG, ParMeTiS, and PT-SCOTCH are presented. Test results on Jüelich BlueGene/P cluster are presented.

## 1. Introduction

The dynamics of near Earth space environment have gained immense importance since many mission critical global technological systems depend on spacecraft that traverse this space and even small dynamical events can cause failures on the functionalities of these spacecraft. Hence performing accurate space weather forecasts are of topmost importance. Space weather forecasting is performed by modeling the electromagnetic plasma system within the near Earth space including the ionosphere, magnetosphere, and beyond.

Vlasiator is a hybrid-Vlasov simulation code developed at the Finnish Meteorological Institute (FMI) and it can be used for very accurate space weather forecasting. In a Vlasov model electrons are fluid and ions are distribution functions, enabling the description of multi-component plasmas without noise. In this model, every grid cell of the ordinary space contains a velocity space, making the simulation six-dimensional. A large-scale Vlasov-hybrid simulation is highly challenging since realistic simulations need to be executed in $\sim 10^6$ spatial grid cells for $\sim 10^6$ time-steps indicating petascale computing.

In Vlasiator the ions are described by a six-dimensional density distribution function. The density distribution function is modeled with a three-dimensional spatial grid and each spatial grid cell contains a three-dimensional velocity distribution function, which is implemented as a simple block-structured grid. For determining the distribution of spatial grid to processors Vlasiator uses the PHG partitioning mode of the Zoltan partitioning framework.

The achievements reported in this whitepaper can be listed as follows:

- Porting of the Vlasiator code to Jüelich BlueGene/P (Jugene) system is done.
- Profiling of the performance of the Vlasiator code up to $10^4$ cores (previous tests were performed for less than $10^3$ cores due to limited resources) is achieved to reveal that Vlasiator successfully scales to $10^4$ cores.
- Analysis of the load-balancing (partitioning) scheme utilized in Vlasiator is performed and it is observed that the time spent on preprocessing constitutes a significant portion of the overall runtime.
- Alternative load-balancing schemes (based on graph-partitioning), which are known to have lower preprocessing costs, are proposed, and embedded in Vlasiator. Experiments show that at least one of the alternative schemes has runtime performances that are as good as PHG.

## 2. Work Done

To investigate the scalability of the Vlasiator code and the effect of preprocessing (partitioning) step in its performance, we ported the Vlasiator code in Jugene. Furthermore, we modified the Zoltan partitioning framework [1] in Vlasiator such that it can now also make use of the parallel graph partitioning tools ParMeTiS [2], and PT-SCOTCH [3] for domain partitioning.

We should note here that graph partitioning models cannot exactly model the communication overheads associated with the communication patterns in Vlasiator. However, as a general observation we can state that, although the communication metrics optimized by graph partitioning schemes are not exact, if the problem domain is regular enough, the error made by graph partitioning method for estimating the communication overhead of a partition is more or less the same for all possible partitions in the solutions space. This property enables the graph partitioning schemes to improve its solutions over regular computational domains successfully since the error made while moving through different partitions in the solution space cancel each other. Since the subject problem domain exhibits such features, we believe that the usage of graph partitioning tools for Vlasiator will yield good results as well.

We compared the performance of the three parallel partitioning tools, namely Zoltan parallel hypergraph partitioner (PHG), ParMeTiS, and PT-SCOTCH, in terms of preprocessing (partitioning) overhead, the communication overhead of the obtained partitions and the load-balancing quality up to 4K cores on Jugene. We called ParMeTiS and PT-SCOTCH from within the Zoltan framework so as to not change the interface of the code. We plan to extend the number of processors/cores used in the experiments to be able to assess the scalability of the partitioning tools further. The metrics we considered for balancing quality involves computational load-balance, total communication volume and communication load-balance.

## 3. Results Obtained

In our experiments we measured the weak and strong scaling performance of Vlasiator while utilizing PHG, ParMeTiS, and PT-SCOTCH for partitioning. In the experiments for weak scaling, 3D grid size is arranged such that under perfect load balance, each process would have to process eight spatial cells, and for strong scaling, total number of spatial cells is set to 32x32x16.

Fig. 1 shows the effects of the partitioning tools to the overall runtime performance of Vlasiator. As seen in the figure for weak scaling experiments (Fig. 1(a)), the overall runtime of Vlasiator (including the preprocessing step) mostly remains in the band between 40 and 60 seconds regardless of the partitioning tool utilized. However a small increase in runtime with increasing number of cores can be observed. As seen in the figure for strong scaling experiments (Fig. 1(b)), regardless of the partitioning tool utilized, when the number of cores double, the overall runtime reduces by a quarter. The reason of this scalability problem is related due to the time spent in preprocessing step as will be seen in Fig. 2. Among the three tools, PT-SCOTCH seems to yield slightly better overall runtimes.

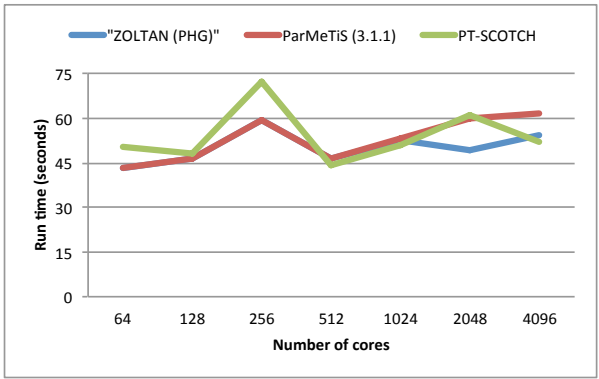 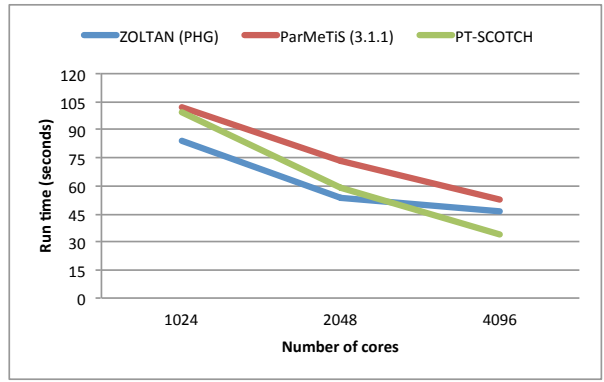

(a) Weak Scaling  (b) Strong Scaling

Fig. 1. Runtime performance of Vlasiator utilizing different partitioning tools.

Figure 2 shows the time spent in preprocessing within the weak and strong scaling experiments while utilizing the domain partitioning tools. We note that the time spent in partitioning increases dramatically with increasing number of processors and starts to constitute a significant portion of the overall runtime. We can observe that all three tools perform in a similar way. This may seem awkward at first since graph-partitioning tools are known to perform much faster than hypergraph-partitioning tools. However, as mentioned before, this equality in performance is due to the fact that all partitioning tools are called from within the Zoltan framework. This means that for running the GP tools, we first convert all data structures to Zoltan specific data structures, then Zoltan converts them back to data structures supported by the GP tools, partitions, and converts back the results to Zoltan specific data structures. This extra overhead can be removed by embedding GP support within the Vlasiator code and then the runtimes obtained via GP tools would be more prominent.

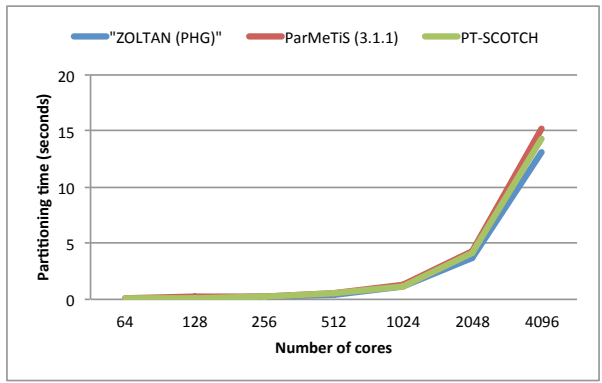 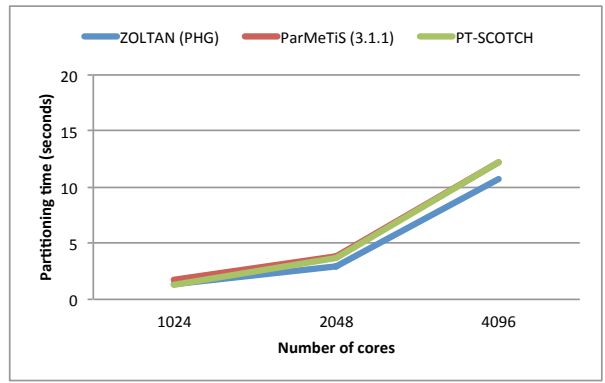

(a) Weak Scaling  (b) Strong Scaling

Fig. 2. Preprocessing time overheads of partitioning tools.

Figure 3 shows the computational imbalance values obtained by running the Vlasiator code and computing the time spent on each processor. Among the partitioning tools, PT-SCOTCH seems to perform the best both for weak and strong scaling experiments, which strengthens the expectation that PT-SCOTCH will perform better in further experiments with larger number of cores.

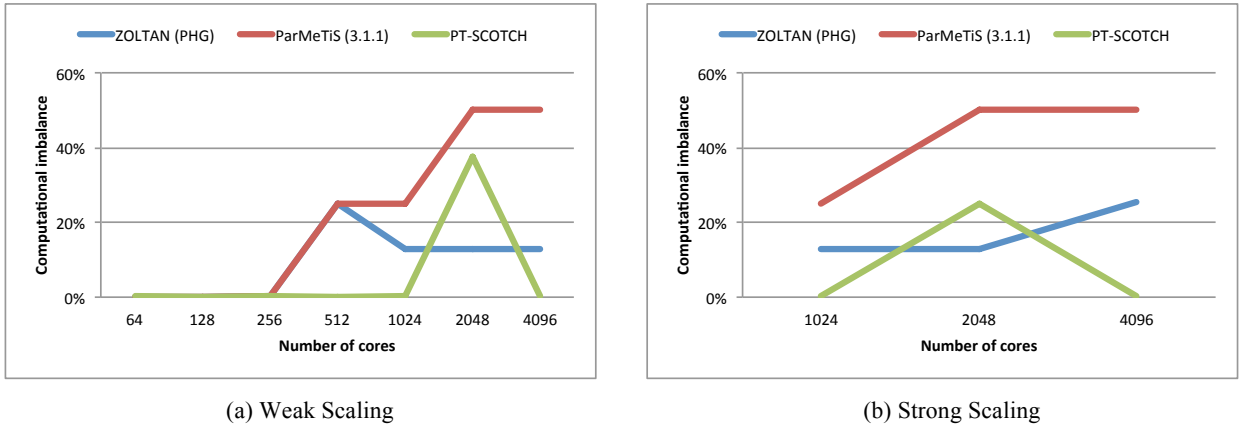

(a) Weak Scaling  (b) Strong Scaling

Fig. 3. Computational imbalance values for Vlasiator utilizing different partitioning tools.

## 4. Conclusion

Collected experimental results indicate that the parallel graph partitioning tools perform almost as good as the parallel hypergraph partitioning tool. Thus, considering their lower running time, we suggest modification of Vlasiator such that it directly utilizes parallel graph partitioning tools. As a future work, we will extend our investigations to include version 4.0.2 of ParMeTiS. We also plan on redesigning the Vlasiator code such that graph partitioning support will be native.

**Acknowledgements**

This work was financially supported by the PRACE project funded in part by the EUs 7th Framework Programme (FP7/2007-2013) under grant agreement no. RI-211528 and FP7-261557. The work is achieved using the PRACE Research Infrastructure resources [give the machine names, and the corresponding sites and countries].

**References**

1. K. Devine, E. Boman, R. Heaphy, B. Hendrickson, and Courtenay Vaughan, Zoltan Data Management Services for Parallel Dynamic Applications, Computing in Science and Engineering, vol. 4(2), pp.: 90—97, 2002.
2. K. Schloegel, G. Karypis, and V. Kumar, Parallel static and dynamic multi-constraint graph partitioning, Concurrency and Computation: Practice and Experience, vol.: 14(3), pp.: 219—240, 2002.
3. C. Chevalier, F. Pellegrini, PT-Scotch: A tool for efficient parallel graph ordering, Parallel Computing, vol.: 34(6-8), pp.: 318-331, 2008.